\documentclass{amsart}

\usepackage{amssymb}

\usepackage{amsthm}

\usepackage{amsfonts}
\usepackage{amsmath}

\def\RR{{\mathbb{R}}}

\def\NN{{\mathbb{N}}}
\def\PP{{\mathbb{P}}}
\def\KK{{\mathbb{K}}}
\def\FF{{\mathbb{F}}}
\def\PP{{\mathbb{P}}}

\def\LL{{\mathbb{L}}}

\def\Spect{{\mathrm{Spect}}}

\def\Ac{\mathcal{U}}

\def\bigO{\mathcal{O}}
\def\bigOsoft{\tilde{\mathcal{O}}}

\def\trdeg{{\textrm{tr.deg}}}

\newtheorem{Thm}{Theorem}
\newtheorem{Lem}[Thm]{Lemma}
\newtheorem{Prop}[Thm]{Proposition}
\newtheorem{Cor}[Thm]{Corollary}

\theoremstyle{definition}
\newtheorem{Def}[Thm]{Definition}

\theoremstyle{remark}
\newtheorem{Rem}[Thm]{Remark}

\theoremstyle{remark}
\newtheorem{Ex}[Thm]{Exemple}

\begin{document}

\title[Decomposition of Multivariate rational Functions]{Nearly Optimal Algorithms for the\\ Decomposition of Multivariate Rational Functions and the Extended L\"uroth's Theorem}

\author[G.~Ch\`eze]{Guillaume Ch\`eze}
\address{Institut de Math\'ematiques de Toulouse\\
Universit\'e Paul Sabatier Toulouse 3 \\
MIP B\^at 1R3\\
31 062 TOULOUSE cedex 9, France}
\email{guillaume.cheze@math.univ-toulouse.fr}

\begin{abstract}
The extended L\"uroth's Theorem says that if the transcendence degree of $\KK(\mathsf{f}_1,\dots,\mathsf{f}_m)/\KK$ is 1 then there exists $f \in \KK(\underline{X})$ such that $\KK(\mathsf{f}_1,\dots,\mathsf{f}_m)$ is equal to $\KK(f)$. In this paper we show how to compute $f$ with a probabilistic algorithm. We also describe a probabilistic and a deterministic algorithm for the decomposition of multivariate rational functions. The probabilistic algorithms proposed in this paper  are softly optimal when $n$ is fixed and $d$ tends to infinity. We also give an indecomposability test based on gcd computations and  Newton's polytope. In the last section, we show that we  get a polynomial time algorithm,  with a minor modification in the exponential time decomposition algorithm proposed by Gutierez-Rubio-Sevilla in 2001.
\end{abstract}

\maketitle


\section*{Introduction}

Polynomial decomposition is the problem of representing a given polynomial $f(x)$ as a functional composition $g(h(x))$ of polynomials of smaller degree. This decomposition has been widely studied since 1922, see \cite{Ritt}, and efficient algorithms are known in the univariate case, see \cite{Alagar_Thanh, Barton_Zippel, Kozen_Landau, Gathen_tame, Gathen_wild} and in the multivariate case \cite{Dickerson, Gathen_tame,GatGutRub}. 

\medskip

The decomposition of rational functions has also been studied,  \cite{Zippel, Alonso_Gut_Recio}. In the multivariate case the situation is the following:\\
Let $f(X_1,\dots, X_n)=f_1(X_1,\dots,X_n)/f_2(X_1,\dots,X_n) \in \KK(X_1,\dots,X_n)$ be a rational function, where $\KK$ is a field and $n\geq 2$. It is commonly said to be composite if it can be written $f=u \circ h$ where $h(X_1,\dots,X_n) \in \KK(X_1,\dots,X_n)$ and $u \in\KK(T)$ such that $\deg(u) \geq 2$ (recall that the degree of a rational function is the maximum of the degrees of its numerator and denominator after reduction), otherwise $f$ is said to be non-composite.

\medskip

This decomposition appears when we study the kernel of a derivation, see \cite{Ollagnier}.  In \cite{Ollagnier} the author gives a multivariate rational function decomposition algorithm, but this algorithm is  not optimal and works only  for fields of characteristic zero. In this paper, we give a probabilistic optimal algorithm.  In other words, our algorithm decomposes $f \in \KK(X_1,\dots,X_n)$ with $\bigOsoft(d^n)$ arithmetic operations, where $d$ is the degree of $f$. We suppose in this work that $d$ tends to infinity and $n$ is fixed.  We use the classical $\bigO$ and $\bigOsoft$ (``soft $\bigO$'') notation in the neighborhood of infinity as defined in~\cite[Chapter~25.7]{GG}. Informally speaking, ``soft $\bigO$''s are used for readability in order to hide logarithmic factors in complexity estimates. Then, the size of the input and the number of arithmetic operations performed by our algorithm have the same order of magnitude. This is the reason why we call our algorithm ``optimal''. \\
Furthermore, our algorithm also works  if the characteristic of $\KK$ is greater than $d(d-1)+1$.\\

This decomposition also appears when we study intermediate fields of an unirational field.  In this situation, the problem is the following: we have $m$ multivariate rational functions $\mathsf{f}_1(\underline{X})$, \dots, $\mathsf{f}_m(\underline{X}) \in \KK(\underline{X})$, and we want to know if there exists  a proper intermediate field $\FF$ such that $\KK(\mathsf{f}_1,\dots,\mathsf{f}_m) \subset \FF \subset \KK(\underline{X})$. In the affirmative case, we want to compute $\FF$. If $\trdeg_{\KK} (\FF) =1$ then by the extended L\"uroth's Theorem, see \cite[Theorem 3 p. 15]{Schinzel} we have $\FF=\KK(f)$.

\begin{Thm}[Extended L\"uroth's Theorem]\label{ext_luroth}
Let $\FF$ be a field such that $\KK \subset \FF \subset \KK(X_1,\dots,X_n)$ and $\trdeg_{\KK} (\FF)=1$. Then there exists $f \in \KK(X_1,\dots,X_n)$ such that $\FF=\KK(f)$. 
\end{Thm}

The classical L\"uroth's Theorem is stated with univariate rational functions. Theorem \ref{ext_luroth} gives an extension to multivariate rational functions. This extended theorem was first proved by Gordan  in characteristic zero, see \cite{Gordan}, and by Igusa in general, see \cite{Igusa}. There exist algorithms to compute $f$, called a L\"uroth's generator,  see e.g. \cite{Gut_Rub_Sev, MullerQuade}. \\
Thanks to the Extended L\"uroth's Theorem the computation of intermediate fields is divided into two parts: first  we compute a L\"uroth's generator $f$, and second we  decompose $f$. Then $f=u \circ h$, and $\FF=\KK(h)$  is an intermediate field. In \cite{Gut_Rub_Sev} the authors show that the decomposition of $f$ bijectively corresponds  to intermediate fields. They also give algorithms to compute a L\"uroth's generator and to decompose it. Unfortunately, the decomposition algorithm has an exponential time complexity, but the complexity analysis of this  algorithm is too pessimistic. Indeed, in the last section of this paper we show that we can modify it  and get an algorithm with a polynomial time complexity.\\

The decomposition of rational functions also appears when  we study the spectrum of a rational function.  In this paper we use this point of view in order to give fast algorithms.\\
Let $\overline{\KK}$ be an algebraic closure of $\KK$. Let $f=f_1/f_2 \in \KK(X_1,\dots,X_n)$ be a rational function of degree $d$. The set 
\begin{eqnarray*}
\sigma(f_1,f_2) =\{(\mu:\lambda) \in \PP^1_{\KK} &\mid &\mu f_1- \lambda f_2\textrm{ is reducible in } \overline{\KK}[X_1,\dots,X_n], \\
&&\textrm{ or } \deg (\mu f_1 - \lambda f_2 ) < d \, \}
\end{eqnarray*} 
is the spectrum of $f=f_1/f_2$. We recall that  a polynomial reducible in $\overline{\KK}[X_1,\dots,X_n]$ is said to be absolutely reducible.\\
A classical theorem of Bertini and Krull, see Theorem \ref{Bert-Krull}, implies that $\sigma(f_1,f_2)$ is finite if $f_1/f_2$ is non-composite. Actually, $\sigma(f_1,f_2)$ is finite if and only if $f_1/f_2$ is non-composite and if and only if the pencil of algebraic  curves $\mu f_1 - \lambda f_2 =0$ has an irreducible general element (see for instance \cite[Chapitre 2, Th\'eor\`eme 3.4.6]{Joua_Pfaff} and \cite[Theorem 2.2]{Bodin} for detailed proofs).\\
 To the author's knowledge, the first effective result about the spectrum has been given by Poincar\'e \cite{Poin}. He showed that $ |\sigma(f_1,f_2)| \leq (2d-1)^2+2d+2$.
This bound was improved  by Ruppert \cite{Ruppert} who proved that 
$$|\sigma(f_1,f_2)| \leq d^2-1.$$
 This result was obtained as a byproduct of a very interesting technique developed to decide the reducibility of an algebraic plane curve.\\
Several papers improve this result,  see e.g. \cite{Lo,Vi,AHS,Bodin, Bus_Che}. 

\medskip

The previous result says that if  $f_1/f_2$ is a non-composite reduced rational function then for all but a finite number of $\lambda \in  \KK$ we have: $f_1 + \lambda f_2$ is absolutely irreducible (i.e. irreducible in $\overline{\KK}[X_1,\dots,X_n]$). Furthermore, the number of ``bad'' values of $\lambda$ is lower than $d^2-1$. Thus we can deduce a probabilistic test for the decomposition of a rational function, based on an absolute irreducibility test. In this paper we will give a decomposition algorithm based on this kind of idea. Furthermore, we will see that this algorithm is softly optimal when the following hypotheses are satisfied:
\begin{center}{ Hypothesis (C):\\
  $\KK$ is a perfect field of  characteristic $0$ or at least $d(d-1)+1$.}\\
\end{center}

\medskip

\begin{center}{Hypothesis  (H):}\end{center}
$$\left\{
\begin{array}{l}
(i)  \deg (f_1+\Lambda f_2)=\deg_{X_n}(f_1 + \Lambda f_2), \textrm{ where } \Lambda \textrm{ is a new variable},\\
(ii)  \, Res_{X_n} \Big( f_1(\underline{0},X_n)+\Lambda f_2(\underline{0},X_n), \, \partial_{X_n} f_1(\underline{0},X_n) + \Lambda  \partial_{X_n} f_2(\underline{0},X_n)\Big) \neq 0 \textrm{ in } \KK[\Lambda].
\end{array}
\right.
$$
where $\deg_{X_n} f$ represents the partial degree of $f$ in the variable $X_n$, $\deg f$ is the total degree of $f$ and $Res_{X_n}$ denotes the resultant relatively to the variable $X_n$.

\medskip

These hypotheses are necessary, because we will use the factorization algorithms proposed in \cite{Lec2007}, where these kinds of hypotheses are needed.  Actually, in \cite{Lec2007} the author studies the factorization of a polynomial $F$ and uses hypothesis (C) and hypothesis (L), where (L) is the following:
\begin{center}{ Hypothesis  (L):}\end{center}
$$\left\{
\begin{array}{l}
(i) \deg_{X_n}F=\deg F, \, \textrm{ and } F \textrm{ is monic in } X_n, \\
(ii)  \, Res_{X_n} \big( F(\underline{0},X_n), \frac{\partial F}{\partial X_n}(\underline{0},X_n) \Big) \neq 0.
\end{array}
\right.
$$
If $F$ is squarefree, then  hypothesis (L)  is not restrictive since it  can be assured by means of a generic linear change of variables, but we will not discuss this question here (for a complete treatment in the bivariate case, see \cite[Proposition 1]{CL}).

\medskip

Roughly speaking, our hypothesis (H) is the hypothesis (L) applied to the polynomial $f_1 + \Lambda f_2$. In (H,$i$) we do not assume that $f_1+ \Lambda f_2$ is monic in $X_n$. Indeed, after a generic linear change of coordinates, the leading coefficient relatively to $X_n$ can be written: $a+\Lambda b$, with $a,b \in \KK$. In our probabilistic algorithm, we evaluate $\Lambda$ to $\lambda \not \in \sigma(f_1,f_2)$, thus $\deg (f_1 +\lambda f_2)= \deg( f_1 + \Lambda f_2)$ and $a + \lambda b \neq 0$. Then we can consider the monic part of $f_1+ \lambda f_2$ and we get a polynomial satisfying (L,$i$). Then (H,$i$) is sufficient in our situation. Furthermore, in this paper, we assume $f_1/f_2$ to be  reduced, i.e. $f_1$ and $f_2$ are coprime. We recall in Lemma \ref{Cor_f+gsquarefree} that in this situation $f_1 + \Lambda f_2$ is squarefree. Thus hypothesis (H) is not restrictive.

\subsection*{Complexity model}
In this paper the  complexity estimates charge a constant
cost for each arithmetic operation ($+$, $-$, $\times$, $\div$) and
the equality test. All the constants in the base fields (or rings)  are thought to be freely at our disposal.
\par
In this paper we suppose that \emph{the number of variables $n$ is fixed} and that the degree $d$ tends to infinity. Furthermore, we say that an algorithm is softly optimal if it works with $\bigOsoft(N)$ arithmetic operations where $N$ is the size of the input.
\par
Polynomials are represented by dense vectors of their
coefficients in the usual monomial basis. For each integer $d$, we
assume that we are given a computation tree that computes the product
of two univariate polynomials of degree at most $d$ with at most $\bigOsoft(d)$
 operations, independently of the base ring, see \cite[Theorem~8.23]{GG}.\\
We use the constant $\omega$ to denote a \emph{feasible matrix
multiplication} exponent as defined in~\cite[Chapter~12]{GG}:
two~$n\times n$ matrices over $\KK$ can be multiplied
with~$\bigO(n^\omega)$ field operations. As in~\cite{BiPa1994}
we require that $2 < \omega \le 2.376$. We recall that the computation of a solution basis  of a linear system with $m$ equations and $d\leq m$ unknowns over $\KK$ takes $\bigO(md^{\omega-1})$ operations in $\KK$ \cite[Chapter~2]{BiPa1994} (see
also~\cite[Theorem~2.10]{Storjohann2000}).\\
In \cite{Lec2007} the author gives a probabilistic (resp. deterministic)  algorithm for the multivariate rational factorization. The rational factorization of a polynomial $f$ is the factorization in $\KK[\underline{X}]$, where $\KK$ is  the coefficient field of $f$. This algorithm uses one factorization of a univariate polynomial of degree $d$ and $\bigOsoft(d^n)$ (resp. $\bigOsoft(d^{n+\omega-1})$) arithmetic operations, where $d$ is the total degree of the polynomial and $n \geq 3$ is the number of variables. If $n=2$, in  \cite{Lec2006},\cite[Errata]{Lec2007}, the author gives a probabilistic (resp. deterministic) algorithm for the rational factorization. The number of arithmetic operations of this algorithm belongs to $\bigOsoft(d^3)$ (resp. $\bigOsoft(d^{\omega +1})$). We note that for $n\geq 3$ if the cost of the univariate polynomial factorization belongs to $\bigOsoft(d^n)$ then the probabilistic algorithm is softly optimal.

\subsection*{Main Theorems}
The following theorems give the complexity results about our algorithms. 
Although we will use no probabilistic model of computation, we will
informally say \emph{probabilistic algorithms} when speaking about the
computation trees occurring in the next theorems. For the sake of
precision, we prefer to express the probabilistic aspects in terms of
families of computation trees. Almost all the trees of a family are
expected to be executable on a given input (if the cardinality of $\KK$
is large enough).

\begin{Thm}\label{main_thm}
Let $f=f_1/f_2$ be a multivariate rational function in $\KK(X_1, \dots,X_n)$ of degree $d$, there exists a family of computation trees over $\KK$ parametrized by
$z:=(\underline{a},\underline{b}) \in \KK^{2n}$  such that:
\begin{itemize}
  \item Any executable tree of the family returns a decomposition $u \circ h$ of $f$ with $h$ a non-composite rational function.
  \item If  $\underline{a},\underline{b}$ are not the roots of some non-zero polynomials the tree corresponding to $z$ is executable.
\end{itemize}
Furthermore, we have:
\begin{enumerate}
\item An executable tree performs two factorizations in $\KK[X_1,\dots,X_n]$ of polynomials with degree $d$, and one computation of $u$.
\item Under hypothesis (C) and (H) we have this estimate: an executable tree performs one factorization of a univariate polynomial of degree $d$ over $\KK$ plus a number of operations in $\KK$ belonging to $\bigOsoft(d^n) $ if $n\geq 3$, or to $\bigOsoft(d^{3})$ if $n=2$.
\end{enumerate}
\end{Thm}

Since we use the dense representation of $f_1$ and $f_2$, the size of $f$ is of the order of magnitude $d^n$. The previous statement thus asserts that the complexity of our probabilistic algorithm is softly optimal for $n \geq 3$.\\
We  precise the condition ``If  $\underline{a},\underline{b}$ are not the roots of some non-zero polynomials'' in Remark \ref{remark_poly_sans_hyp_complexity} and Remark \ref{remarksomepoly}.\\
In characteristic zero we can say that for almost all $z$ the  tree corresponding to $z$ is executable.\\

We also give a deterministic decomposition algorithm.

\begin{Thm}\label{maindet}
If $\KK$ is a field with a least $\max(d^2,\frac{3}{2}d^2-2d+1)$ elements, then the decomposition $f=u \circ h$, with $h$ non-composite, can be computed with at most $\bigO(d^2)$ absolute factorizations of  polynomials with degree  $d$, and at most $\bigO(d^2)$ computations of $u$ where $f$ and $h$ are given.
\end{Thm}

If we can use the algorithm proposed in \cite{CL} and \cite{Lec2007}, as we will see in Remark \ref{remalgodet},  our deterministic algorithm uses one factorization of a univariate polynomial of degree $d$  with algebraic coefficients of degree at most $d$, and at most $\bigOsoft(d^{n+\omega+2})$ if $n\geq 3$  or $\bigOsoft(d^6)$ if $n=2$ arithmetic operations in $\KK$. \\

With the tools used for the decomposition algorithms, we can compute a L\"uroth's generator.

\begin{Thm}\label{mainluroth}
Let  $\mathsf{f}_1,\dots,\mathsf{f}_m\in \KK(X_1,\dots,X_n)$ be  $m$ rational functions of degree at most $d$. There exists a family of computation trees over $\KK$ parametrized by $z=(z_1,\dots,z_m) \in \KK^{2nm}$, such that:\\
If for all $i=1,\dots,m$, $z_i \in \KK^{2n}$ belongs to an open Zariski set related to $\mathsf{f}_1, \dots,\mathsf{f}_i$ then the tree corresponding to $z$ is executable on $\mathsf{f}_1,\dots,\mathsf{f}_m$ and it returns a L\"uroth's generator of $\KK(\mathsf{f}_1,\dots,\mathsf{f}_m)$.\\
Furthermore, we have:
\begin{enumerate}
\item An executable tree performs $2m$ gcd computations in $\KK[X_1,\dots,X_n]$ with polynomials of degree at most $d$.
\item If $\KK$ has at least  $(4d+2)d$ elements then we have the estimate: an executable tree performs $\bigOsoft(md^n)$ arithmetic operations in $\KK$. 
\end{enumerate}

\end{Thm}

As before,  this algorithm is softly optimal because the order of magnitude of the input is $md^n$. A precise description  of the open Zariski set is given in Remark \ref{remarkzariskiluroth1}.

\medskip

In the last section we prove the following result:
\begin{Thm}\label{mainschichothm}
Let $f=f_1/f_2 \in \KK(\underline{X})$.\\
$f=u \circ h$, with $h=h_1/h_2$    if and only if $H(\underline{X},\underline{Y})=h_1(\underline{X})h_2(\underline{Y})- h_2(\underline{X})h_1(\underline{Y})$ divides $F(\underline{X},\underline{Y})=f_1(\underline{X})f_2(\underline{Y})- f_2(\underline{X})f_1(\underline{Y})$.\\
Furthermore, if $h_1/h_2$ is a reduced non-composite rational function then $H$ is one of the irreducible factors with the smallest degree relatively to $\underline{X}$ of $F$.
\end{Thm}

The first part of this theorem is already known, see \cite{Schicho}. Here, we prove that $H$ is irreducible if $h_1/h_2$ is non-composite. This result implies that we can modify the exponential time decomposition algorithm presented in \cite{Gut_Rub_Sev} and get a polynomial time algorithm.
\subsection*{Comparison with other algorithms}
There already exist several algorithms for the decomposition of rational functions. In \cite{Gut_Rub_Sev}, the authors provide two algorithms to decompose a multivariate rational function. These algorithms run in exponential time in the worst case. In the first one we have to factorize \mbox{$f_1(\underline{X})f_2(\underline{Y})-f_1(\underline{Y})f_2(\underline{X})$} and to look for factors of the following kind $h_1(\underline{X})h_2(\underline{Y})-h_1(\underline{Y})h_2(\underline{X})$. The authors say that in the worst case the number of candidates to be tested is exponential in $d=\deg( f_1/f_2)$. In the last section we show that actually the number of candidates is bounded by $d$. Thus we can get a polynomial time algorithm.\\
  In the second algorithm, for each pair of factors $(h_1,h_2)$ of $f_1$ and $f_2$ (i.e. $h_1$ divides $f_1$ and $h_2$ divides $f_2$), we have to test if there exists $u \in \KK(T)$ such that $f_1/f_2=u(h_1/h_2)$. Thus in the worst case we also have an exponential number of candidates to be tested.\\
To the author's knowledge, the first polynomial time algorithm is due to J.-M. Ollagnier, see \cite{Ollagnier} . This algorithm relies on the study of the kernel of the following derivation: $\delta_{\omega}(F)=\omega \wedge dF$, where $F \in \KK[\underline{X}]$ and $\omega=f_2df_1-f_1df_2$. In \cite{Ollagnier} the author shows that we can reduce the  decomposition of a rational function to linear algebra.  The bottleneck of this algorithm is the computation of the kernel of a matrix. The size of this matrix is $\bigO(d^n) \times \bigO(d^n)$, then the complexity of this deterministic algorithm belongs to $\bigO(d^{n\omega})$.  In  \cite{Ollagnier}, as in this paper, the study of the pencil $\mu f_1 - \lambda f_2$ plays a crucial role.

\subsection*{Structure of this paper} In Section 1, we give a toolbox where we recall some results about decomposition and factorization. In Section 2, we describe our algorithms to decompose multivariate rational functions. In Section 3, we give an indecomposability test based on the study of a Newton's polytope. In Section 4, we give two algorithms to compute a L\"uroth's generator. In Section 5 we show that the decomposition algorithm presented in \cite{Gut_Rub_Sev} can be modified to get a polynomial time complexity algorithm.

\subsection*{Notations}
All the rational functions are supposed to be reduced.\\
Given a polynomial $f$, $\deg(f)$ denotes its total degree.\\
$\overline{\KK}$ is an algebraic closure of $\KK$.\\
For the sake of simplicity, sometimes we write $\KK[\underline{X}]$ instead of $\KK[X_1,\dots,X_n]$, for $n\geq 2$.\\
$Res(A,B)$ denotes the resultant of two univariate polynomials $A$ and $B$.\\
For any polynomial $P \in \overline{\KK}[\underline{X}]$, we write $\mathcal{U}(P):= \{ a \in \KK^n \mid P(a) \neq 0 \}$. \\

\section{Prerequisite}\label{toolbox}


  The following result implies, as mentioned in the introduction, that hypothesis (H) is not restrictive.

\begin{Lem}\label{Cor_f+gsquarefree}
If $f_1/f_2$ is reduced in $\KK(X_1,\dots,X_n)$, where $n\geq 1$ and $\Lambda$ is a  variable, then $f_1+\Lambda f_2$ is squarefree.
\end{Lem}

Now we introduce our main tools.

\begin{Prop}\label{Thm_Bodin} Let $f=f_1/f_2$ be a  rational function in $\KK(X_1,\dots,X_n)$.\\
$f$ is composite if and only if  $\mu f_1-\lambda f_2$ is reducible in $\overline{\KK}[\underline{X}]$ for all $\mu, \lambda \in \overline{\KK}$ such that $\deg (\mu f_1 - \lambda f_2)= \deg (f)$.\\
We also have: $f$ is non-composite if and only if its spectrum $\sigma(f_1,f_2)$ is finite, \\if and only if $f_1-T f_2$ is absolutely irreducible in $\overline{\KK(T)}[\underline{X}]$, where $T$ is a new variable.\\
Furthermore if $\deg (f) =d$ then $\sigma(f_1,f_2)$ contains at most $d^2-1$ elements.
\end{Prop}
\begin{proof}
The first part of this result was known by Poincar\'e see \cite{Poin}, for a modern statement and a proof, see \cite[Corollary 2.3]{Bodin}.\\
 The  bound $|\sigma(h_1,h_2)|\leq d^2-1$ is proved for any field in the bivariate case in  \cite{Lo}. We deduce the multivariate case easily thanks to the Bertini's irreducibility theorem, see e.g. \cite{Bodin} or the proof of Theorem 13 in \cite{Bus_Che} for an application of the Bertini's irreducibility theorem in this context.\\
\end{proof}

\begin{Lem}\label{Lem_Bodin}
Let $h=h_1/h_2$ be a rational function in $\KK(\underline{X})$, $u=u_1/u_2$ a  rational function in $\KK(T)$ and set $f=u \circ h$ with $f=f_1/f_2 \in \KK(\underline{X})$. 
For all $\lambda \in \KK$ such that $\deg (u_1 - \lambda u_2) = \deg u$, we have
$$f_1 - \lambda f_2= e (h_1 - t_1 h_2)\cdots (h_1 -t_k h_2)$$
where $e \in \KK$, $k= \deg u$ and $t_i  \in \overline{\KK}$ are the roots of the univariate polynomial $u_1(T) - \lambda u_2(T)$. 
\end{Lem}
\begin{proof}
See the proof of Lemma \ref{Lem_Bodin_gen} in Section \ref{appendix}. Lemma \ref{Lem_Bodin_gen} is a generalization of Lemma \ref{Lem_Bodin}. We state Lemma \ref{Lem_Bodin} in our toolbox because the generalization will be only used  in Section \ref{appendix}.
\end{proof}

\begin{Rem}
If $t_i \in \KK$ then  $h_1 - t_i h_2 \in \KK[X_1,\dots,X_n]$ is an irreducible factor of $f_1 - \lambda f_2$. Thus with a rational factorization we get  information about the decomposition of $f$. This remark will be used during our probabilistic decomposition algorithm in order to avoid an absolute factorization.
\end{Rem}



\section{Decomposition algorithms}\label{algo_prob}

\subsection{Computation of  $u$}\label{sec_comput_u}
Suppose that $f=f_1/f_2=u \circ h \in \KK(X_1,\dots,X_n)$, $h \in \KK(X_1,\dots,X_n)$, and $u \in \KK(T)$. We set $h=h_1/h_2$.\\
Usually, when $h_1$ and $h_2$ are given we get $u=u_1/u_2$ by solving a linear system, see \cite[Corollary 2]{Gut_Rub_Sev}.  Let $\mathcal{M}(h_1,h_2)$ be the matrix corresponding to this linear system in the monomial basis. In our situation the size of $\mathcal{M}(h_1,h_2)$ is $\bigO(d^n)\times\bigO(d)$. Thus we can find $u$ with $\bigOsoft(d^{n+\omega-1})$ operations in $\KK$.\\

We can get $u$ with another approach. This approach is based on a  strategy due to Zippel in \cite{Zippel}. Zippel showed in the univariate case  that we can compute $u$ quickly. His strategy is the following: compute the power series $H$ such that $h\circ H(X)=X$, then compute $f\circ H$, and finally deduce $u$ with a Pad\'e approximant. All these steps can be done with $\bigOsoft(d)$ or $\bigOsoft(d^{3/2})$ arithmetic operations, see \cite[Chapter 1]{BiPa1994}, and \cite{BrentKung}. Thus we deduce that in the univariate case, $u$ can be computed with $\bigOsoft(d^{3/2})$ arithmetic operations.\\
In the multivariate case with hypothesis (H), we have $\deg(f)=\deg_{X_n}(f)$. Thus $f(\underline{0},X_n)=u\circ h(\underline{0},X_n)$ is not a constant. Then we can apply Zippel's strategy to $f(\underline{0},X_n)$ in order to find $u$. This method is correct because if  $f$ and $h$ are given then there is a unique $u$ such that $f= u \circ h$, see \cite[Corollary 2]{Gut_Rub_Sev}. Thus we have proved the following result:
\begin{Lem}\label{lem_comput_u}
Let $f, h \in \KK(X_1,\dots,X_n)$ be rational functions. We suppose that $f$ satisfies hypothesis (H) and we set $\deg(f)=d$. If there exists $u \in \KK(T)$ such that $f=u \circ h$ then we can compute $u$ with $\bigOsoft(d^{n})$ arithmetic operations.
\end{Lem} 
\begin{proof}
We compute  $f(\underline{0},X_n)$ with $\bigOsoft(d^n)$ arithmetic operations. Then we compute $u$ as explained above with $\bigOsoft(d^{3/2})$ arithmetic operations.
\end{proof}

\subsection{A probabilistic algorithm}

\textbf{\textsf{Decomp}}\\
\textbf{Input:} $f=f_1/f_2 \in \KK(X_1,\dots,X_n)$, 
 $z:=(\underline{a},\underline{b}) \in \KK^{2n}$.\\
\textbf{Output:} A decomposition of $f$ if it exists, with $f=u \circ h$, $u=u_1/u_2$, $h=h_1/h_2$ non-composite and $\deg u \geq 2$.\\

\begin{enumerate}
\item We set $F_a=f_2(\underline{a})f_1(\underline{X})-f_1(\underline{a})f_2(\underline{X})$, $F_b=f_2(\underline{b})f_1(\underline{X})-f_1(\underline{b})f_2(\underline{X})$.
\item \label{stepfacto} Factorize $F_a$ and $F_b$.
\item If $F_a$ or $F_b$ is irreducible then Return ``$r$ is non-composite''.
\item \label{stepfactoh}Let $\mathcal{F}_a$ (resp. $\mathcal{F}_b$) be an irreducible factor of $F_a$ (resp. $F_b$) with the smallest degree.
\item Set $h=\mathcal{F}_a/\mathcal{F}_b$.

\item Compute $u$ such that $f=u \circ h$ as explained in Section \ref{sec_comput_u}.
\item Return $u, h$.
\end{enumerate}

\begin{Ex}$\,$
\begin{enumerate}
\item[a-]
We consider $f=f_1/f_2$ , with $f_1=X^3+Y^3+1$ and $f_2=3XY$. We set $\underline{a}=(0,0)$, $\underline{b}=(0,1)$. Then $F_a=-3XY$ and $F_b=3X^3+3Y^3-6XY+3$. $F_a$ is reducible but $F_b$ is irreducible then we conclude that $f$ is non-composite.
\item[b-] Now, we apply the algorithm \textsf{Decomp} to the rational function $f=u \circ h$, where $u=(T^2+1)/T$ and $h=h_1/h_2$ with $h_1= X^3+Y^3+1$ and $h_2=3XY$. We have seen above that $h$ is non-composite.\\
In  this situation with $\underline{a}=(0,0)$ and $\underline{b}=(0,1)$ we get:
$$F_a=-3.X.Y . (X^3+Y^3+1), \textrm{ and }$$
$$F_b=-12. X .Y . (X^3+Y^3 +1).$$
Then the algorithm cannot give a correct output in this situation. Here, we have $f_2(\underline{a})=f_2(\underline{b})$, we will see that we must avoid this situation.\\
If we set $\underline{a}=(2,1)$ and $\underline{b}=(1,-1)$ then:
$$F_a=60.(X^3+Y^3-5XY+1).(X^3+Y^3-\dfrac{3}{5}XY+1), \textrm{ and}$$
$$F_b=-3.(X^3+Y^3+XY+1).(X^3+Y^3+3XY+1).$$
Then we get $\mathcal{F}_a=X^3+Y^3-5XY+1$ and $\mathcal{F}_b=X^3+Y^3+XY+1$. The algorithm \textsf{Decomp} returns $h=\mathcal{F}_a/\mathcal{F}_b$. This is a correct output since $U \circ \mathcal{F}_a/\mathcal{F}_b=h_1/h_2$, where $U=\big(T/6+5/6\big)/\big(-T/2 +1/2\big)$.
\end{enumerate}
\end{Ex}

\begin{Prop}\label{prop_correct}
If $\underline{a}, \underline{b}$ are not the roots of some non-zero polynomials then the algorithm corresponding to $z=(\underline{a},\underline{b})$ is correct.
\end{Prop}

\begin{proof}
First, we suppose that $f$ is non-composite and we set
 $$\Spect_{f_1,f_2}(T_1,T_2)= \prod_{(\mu:\lambda) \in \sigma(f_1,f_2)} (\mu T_2 - \lambda T_1).$$
We have  $\Spect_{f_1,f_2}(\mu, \lambda)=0$ if and only if $(\mu:\lambda) \in \sigma(f_1,f_2)$.\\
If  $\Spect_{f_1,f_2}\big( f_2(\underline{a}),f_1(\underline{a}) \big). \Spect_{f_1,f_2}\big( f_2(\underline{b}),f_1(\underline{b}) \big) \ne 0$ then $F_a$ and $F_b$ are absolutely irreducible and $\deg F_a=\deg F_b=\deg f$.\\
This gives: if $\underline{a}$ and $\underline{b}$ avoid the roots of 
$$S(\underline{A}, \underline{B}):=\Spect_{f_1,f_2}\big( f_2(\underline{A}),f_1(\underline{A}) \big). \Spect_{f_1,f_2}\big( f_2(\underline{B}),f_1(\underline{B}) \big), $$
where $\deg S \leq 2d(d^2-1)$ by Proposition \ref{Thm_Bodin}, then the algorithm returns: ``$r$ is non-composite''.\\

Second, we suppose $f=v \circ H$, with $H \in \KK(X_1,\dots,X_n)$ a non-composite rational function. We set $v=v_1/v_2$, $H=H_1/H_2$ such that these two rational functions  are reduced. We also suppose that $f_2(\underline{a})$ and $f_2(\underline{b})$ are nonzero.\\
If $\deg F_a=\deg F_b=\deg f$ then $\underline{a}$ and $\underline{b}$ are not the roots of a polynomial $D$ of degree $d$. Thanks to Lemma \ref{Lem_Bodin} we have: 
$$F_a=e(H_1- t_1 H_2) \cdots (H_1 -t_k H_2),$$
$$F_b=e'(H_1- s_1 H_2) \cdots (H_1 -s_k H_2),$$
with $e,\, e' \in \KK$, $t_i,\, s_j \in \overline{\KK}$.\\
As $H_1(\underline{a})/H_2(\underline{a})$ (resp. $H_1(\underline{b})/H_2(\underline{b})$) is a root of $f_2(\underline{a})v_1(T)-f_1(\underline{a})v_2(T)$ (resp. $f_2(\underline{b})v_1(T)-f_1(\underline{b})v_2(T)$), we set $t_1=H_1(\underline{a})/H_2(\underline{a})$ and $s_1=H_1(\underline{b})/H_2(\underline{b})$, and we remark that $t_1, s_1 \in \KK$. We set
$$\Spect_{H_1,H_2}(T)= \prod_{\lambda \in \sigma(H_1,H_2) \cap \KK} ( T- \lambda ).$$
If $ \Spect_{H_1,H_2} (t_1) \neq 0$ (resp. $\Spect_{H_1,H_2}(s_1)\neq 0$) then $H_1 - t_1 H_2$ (resp. $H_1 -s_1 H_2$) is absolutely irreducible.\\
If $$R(\underline{a}, \underline{b})=Res_T\big( f_2(\underline{a})v_1(T)-f_1(\underline{a})v_2(T),\, f_2(\underline{b})v_1(T)-f_1(\underline{b})v_2(T) \big) \neq 0$$
 then $t_i \neq s_j$ for all $i,j$. We remark that $R$ is a nonzero polynomial by Lemma \ref{Cor_f+gsquarefree} since $v_1$ and $v_2$ are coprime. Thus step \ref{stepfactoh} gives $\mathcal{F}_a=H_1-tH_2$, $\mathcal{F}_b=H_1 -s H_2$ with $t,s \in \KK$ and $t \neq s$. Then $h=\mathcal{F}_a/ \mathcal{F}_b$ is non-composite, because $H_1/H_2$ is non-composite.
\end{proof}

\begin{Rem}\label{remark_poly_sans_hyp_complexity}
Now, with the notations of the previous proof, we can explain in details the meaning of:
\emph{``If  $\underline{a},\underline{b}$ are not the roots of some non-zero polynomials''} in Proposition \ref{prop_correct} and Theorem \ref{main_thm}. This means:\\
If $f$ is non-composite then there exists a nonzero polynomial 
$$P(\underline{A},\underline{B}):=S(\underline{A},\underline{B})$$ 
of degree at most $2d(d^2-1)$ such that for any $(\underline{a}, \underline{b}) \in \Ac(P)$ the algorithm  corresponding to $z$ is executable and returns a correct output.\\
If $f$ is composite then there exists a nonzero polynomial 
$$D_1(\underline{A},\underline{B}):=f_2(\underline{A}).f_2(\underline{B}).D(\underline{A}).D(\underline{B})$$
 of degree at most $4d$ such that;\\
for any $(\underline{a},\underline{b}) \in \Ac(D_1)$, there exist  nonzero polynomials 
$$D_2(\underline{A}):=\prod_{\lambda \in \sigma(H_1,H_2)\cap \KK}\big(H_2(\underline{A}) - \lambda H_1(\underline{A})\big)$$
 of degree at most $(d^2-1).d/2$, and 
$$R(\underline{A}, \underline{B})$$
where $\deg_{\underline{A}} R \leq d^2/2$ and $\deg_{\underline{B}} R \leq d^2/2$,
such that; 
for any $(\underline{a}, \underline{b}) \in \Ac\big(D_2(\underline{A}).D_2(\underline{B}).R(\underline{A}, \underline{B})\big)$, 
the algorithm corresponding to $z=(\underline{a},\underline{b})$ is executable and returns a correct output.\\
\end{Rem}

\begin{Prop}\label{prop_complexity}
Under hypotheses (C) and (H), if $\underline{a}$ and $\underline{b}$ are not the roots of a non-zero polynomial then we can use the algorithm proposed in \cite{Lec2007}. Then the algorithm \textsf{Decomp} performs one factorization of a univariate polynomial of degree $d$ over $\KK$ plus a number of operations in $\KK$ belonging to $\bigOsoft(d^n) $ if $n\geq 3$ or to $\bigOsoft(d^{3})$ if $n=2$.
\end{Prop}

\begin{proof}
 As $f$ satisfies (H,$i$), we deduce that if $\underline{a}$ and $\underline{b}$ are not the roots of a polynomial $D$ of degree $d$, then the monic part relatively to $X_n$ of $F_a$ (resp. $F_b$) satisfies (L,$i$).\\
We set:
$$\mathcal{D}(\Lambda)=Res_{X_n}\big( f_1(\underline{0},X_n)-\Lambda f_2(\underline{0},X_n),\,  \partial_{X_n} f_1(\underline{0},X_n)-\Lambda \partial_{X_n} f_2(\underline{0},X_n) \big).$$
By hypothesis (H,$ii$), $\mathcal{D}(\Lambda) \neq 0$ in $\KK[\Lambda]$. Furthermore if $f_2(\underline{a})$ and $f_2(\underline{b})$ are nonzero and $\mathcal{D}\big(f_1(\underline{a})/f_2(\underline{a})\big) \neq 0$ (resp. $\mathcal{D}\big(f_1(\underline{b})/f_2(\underline{b})\big) \neq 0$) then hypothesis (L,$ii$) is satisfied for $F_a$ (resp. $F_b$).
Then we can  use Lecerf's algorithm, see \cite{Lec2007}. This gives: if $\underline{a}$ and $\underline{b}$ avoid the roots of 
$$\overline{\mathcal{D}}(\underline{A}, \underline{B})=
\mathcal{D}\big(f_1(\underline{A})/f_2(\underline{A})\big).\mathcal{D}\big(f_1(\underline{B})/f_2(\underline{B})\big).\big(f_2(\underline{A}).f_2(\underline{B})\big)^{\deg \mathcal{D}+1},$$
 and $\deg \overline{\mathcal{D}} \leq 2\big(d(d-1)d+d\big)$ then we can use the algorithm proposed by G.~Lecerf in \cite{Lec2007}.\\
The complexity result comes from  Lemma \ref{lem_comput_u}, and \cite[Proposition 5]{Lec2007}, \cite[Proposition 2]{Lec2006} and \cite[Errata]{Lec2007}.
\end{proof}

\begin{Rem}\label{remarksomepoly}
The meaning of the condition \emph{``if $\underline{a}$ and $\underline{b}$ are not the roots of a non-zero polynomial''} in Proposition \ref{prop_complexity} is the following:  If we want to use Lecerf's factorization algorithm in order to get the complexity estimate given in the second part of Theorem \ref{main_thm}, then $\underline{a}$ and $\underline{b}$ must also avoid the roots of the polynomial $$D(\underline{A}).D(\underline{B}).\overline{\mathcal{D}}(\underline{A}, \underline{B}),$$ where $\deg D \leq d$ and $\deg \overline{\mathcal{D}} \leq 2(d^2(d-1)+d)$.

\end{Rem}

It follows that Theorem \ref{main_thm} comes from Proposition \ref{prop_correct} and Proposition \ref{prop_complexity}.
\subsection{A deterministic algorithm}
\textbf{\textsf{Decomp Det}}\\
\textbf{Input:} $f=f_1/f_2 \in \KK(X_1,\dots,X_n)$, $S=\{s_0,\dots,s_{\mathcal{B}}\}$ a subset  of $\KK$ with at least $\mathcal{B}+1=\max(d^2,\frac{3}{2}d^2-2d+1)$ distinct elements.\\
 \textbf{Output:} A decomposition of $f$ if it exists, with $f=u \circ h$, $u=u_1/u_2$, $h=h_1/h_2$ non-composite and $\deg u \geq 2$.\\

\noindent t:=false, $\lambda:=0$.\\
While t=false do
\begin{enumerate}
\item \label{step1deth} If $\deg (f_1 + s_{\lambda} f_2)=\deg (f)$ then go to step \ref{step2deth} else $\lambda:=\lambda +1$.
\item  \label{step2deth} Compute the absolute factorization of $F_{\lambda}:=f_1+ s_{\lambda} f_2$.
\item \label{stepFabsirred}If $F_{\lambda}$ is absolutely irreducible then Return ``$f$ is non-composite''.
\item If $F_{\lambda}$ is absolutely reducible then
\begin{enumerate}
\item  \label{steph1h2det} If two distinct absolute irreducible factors $\mathsf{f}_1$,  $\mathsf{f}_2$ belong to $\KK[\underline{X}]$ then we set $h_1:=\mathsf{f}_1$ and $h_2:=\mathsf{f}_2$,\\
If there exists an absolute irreducible factor $\mathsf{f}_1:=\mathcal{F}_1 + \epsilon \mathcal{F}_2$, with $\epsilon \in \overline{\KK} \setminus \KK$ and $\mathcal{F}_1, \mathcal{F}_2 \in \KK[\underline{X}]$ then we set $h_1:=\mathcal{F}_1$, $h_2:=\mathcal{F}_2$,\\
Else $\lambda:= \lambda+1$ and go to step \ref{step1deth}.
\item \label{computudet}Compute $u$ (if it exists) such that $f=u\circ h$ as explained in Section \ref{sec_comput_u}.
\item If $u$ exists then t:=true else $\lambda:=\lambda+1$.
\end{enumerate}
\end{enumerate}
Return $u, h$.\\

\begin{Ex}$\,$
\begin{enumerate}
\item[a-] We consider $f=f_1/f_2$, where $f_1=3XY$ and $f_2=X^3+Y^3+1$. This gives $F_0=3.X.Y$, then $F_0$ is reducible, and this gives $h=X/Y$. We do not find a rational function $u$ such that $f=u \circ (X/Y)$ then we consider $F_1=f_1+f_2$. $F_1$ is absolutely irreducible, then the algorithm \textsf{Decomp Det} returns $f$ is non-composite.
\item[b-] Now, we apply the algorithm \textsf{Decomp Det} to the rational function $f=u \circ h$, where $u=(T^2+1)/T$ and $h=(X^3+Y^3+1)/(3XY)$. As we have seen above $h$ is non-composite.\\
In this situation we have:
$$F_0=(X^3+Y^3+1+3.i.X.Y)(X^3+Y^3+1-3.i.X.Y),$$ where $i^2=-1$.\\
Then we have $\mathsf{f}_1=X^3+Y^3+1+3.i.X.Y$, $\mathcal{F}_1=X^3+Y^3+1$, $\mathcal{F}_2=3XY$. The algorithm returns  $\mathcal{F}_1/\mathcal{F}_2=h$.
\end{enumerate}
\end{Ex}

\begin{Prop}\label{detcorrect}
The algorithm is correct. Furthermore we go back to step \ref{step1deth} at most $\bigO(d^2)$ times.
\end{Prop}

\begin{proof}
First, we suppose that $f$ is non-composite. By Proposition \ref{Thm_Bodin} there exists $s_{\lambda_0} \in S$ such that $s_{\lambda_0} \not \in \sigma(f_1,f_2)$ because $S$ contains at least $d^2$ elements. Thus $f_1+s_{\lambda_0} f_2$ is absolutely irreducible and step \ref{stepFabsirred} returns $f$ non-composite.\\
We remark that if $f_1 + s_{\lambda} f_2$ is reducible then we cannot find $u$ during step \ref{computudet} because $f$ is non-composite. Then if $f$ is non-composite the algorithm is correct.

\medskip

Second, we suppose that $f$ is composite and $f=v\circ H$ with  $H=H_1/H_2$ a reduced and non-composite rational function, $\deg v \geq 2$ and $v=v_1/v_2$ is a reduced rational function.\\
$f_1+ s_{\lambda} f_2 = e \prod_i (H_1 +t_i H_2)$ by Lemma \ref{Lem_Bodin}, where $(v_1 + s_{\lambda} v_2)(t_i)=0$. \\
There exists  $s_{\lambda_0} \in S$ such that $D(s_{\lambda_0}) \neq 0$, where 
$$D(\Lambda)=Res(v_1+\Lambda v_2,v'_1+\Lambda v'_2)\times \prod_{x_i \in \sigma(H_1,H_2) \cap \KK} \big( v_2(x_i) - \Lambda v_1(x_i)\big).$$
Indeed $D(\Lambda)$ is a nonzero polynomial by Lemma \ref{Cor_f+gsquarefree} since $v_1$ and $v_2$ are coprime. Furthermore, by Proposition \ref{Thm_Bodin}, we have 
$$\deg D \leq \deg v (\deg v -1) + \big((\deg H)^2 -1\big).\deg v.$$
  As $\deg v. \deg H=d$ and $\deg v \geq 2$ , we get 
$$\deg D \leq 3/2 d^2-2d.$$
 As $S$ contains at least $3/2d^2-2d+1$ distinct elements, there exists $s_{\lambda_0} \in S$  such that  $D(s_{\lambda_0}) \neq 0$ and then for all $i$, $t_i \not \in \sigma(H_1,H_2)$, and $t_i \neq t_j$ for all $i \neq j$. \\
Then for $\lambda_0$ we construct $h_1$ and $h_2$ as explained in step \ref{steph1h2det}. (If $t_1,t_2 \in \KK$  are distinct then we have two absolutely irreducible factors in $\KK[\underline{X}]$, else if $t_1 \in \overline{\KK} \setminus \KK$ then we construct $h_1$ and $h_2$ with only one absolutely irreducible factor.) We have $h_1/h_2=w \circ H_1/H_2$ where $w \in \KK(T)$ and $\deg w=1$.\\
We remark that if $f$ is composite then  we find a decomposition $f=u \circ h$ with $h$  non-composite. Indeed, there exist $(\mu:\lambda)$ and $(\mu':\lambda') \neq (\mu:\lambda) \in \PP^1_{\KK}$ such that $\mu h_1+ \lambda h_2$ and $\mu' h_1+ \lambda' h_2$ are absolutely irreducible. (It is obvious if $t_1,t_2 \in \KK$. If $t_1 \in \overline{\KK}\setminus \KK$ there exists a conjugate $t'_1$ of $t_1$ over $\KK$ such that $h_1 + t'_1h_2$ is absolutely irreducible.) Then $h_1/h_2$ is non composite by Proposition \ref{Thm_Bodin}. Thus if $f$ is non-composite the output is correct.
\end{proof}

Theorem \ref{maindet} is a direct corollary of Proposition \ref{detcorrect}.

\begin{Rem}\label{remalgodet}
In \cite{CL} the authors show that we can compute, under the hypothesis (C), the absolute factorization of a bivariate squarefree polynomial with at most $\bigOsoft(d^4)$ arithmetic operations. As we go back to step 1 at most $\bigO(d^2)$ times we deduce that the algorithm \textsf{Decomp Det} uses at most $\bigOsoft(d^6)$ arithmetic operations.\\
When $n\geq 3$, a complexity analysis of an absolute factorization algorithm  as studied in \cite{CL} is not done, but we can estimate the cost of our deterministic algorithm. Indeed, we can reduce absolute factorization to factorization over a suitable algebraic extension $\KK[\alpha]$ of degree at most $d$ over $\KK$, \cite{Tra76,Tra85,DT89,Kal85}. With this strategy and with the deterministic factorization algorithm proposed in \cite{Lec2007} we get an absolute factorization algorithm which performs at most $\bigOsoft(d^{n+\omega-1})$ arithmetic operations in $\KK[\alpha]$. Thus the algorithm performs $\bigOsoft(d^{n+\omega})$ arithmetic operations in $\KK$, because $[\KK[\alpha]:\KK] \leq d$.  As we go back to step 1 at most  $\bigO(d^2)$ times  we deduce that, if we can use Lecerf's deterministic factorization algorithm, the algorithm \textsf{Decomp Det} uses at most $\bigOsoft(d^{n+\omega+2})$ arithmetic operations and one factorization of a univariate polynomial of degree $d$ with coefficients in $\KK[\alpha]$.\\
\end{Rem}

\section{An indecomposability test using Newton's polytope}\label{sect_indecomp_test}

In Section \ref{algo_prob},  if $f_1$ and $f_2$ are sparse our algorithms do not use this information. 
In this section we give an indecomposability test based on some properties of the Newton's polytope. The idea is to generalize this remark: if $\deg f$ is a prime integer then $f$ is non-composite. This is obvious because $f = u \circ h$ implies $\deg f= \deg u . \deg h$, and $\deg u \geq 2$.
\begin{Def}
Let $f(\underline{X}) \in \KK[X_1,\dots,X_n]$, the support of $f(\underline{X})$ is the  set  $S_f$ of integer points $(i_1,\dots,i_n)$ such that the monomial $X_1^{i_1}\cdots X_n^{i_n}$ appears in $f$ with a nonzero coefficient.\\
We denote by $N(f)$ the convex hull (in the real space $\RR^n$) of
$S_f $.
This set $N(f)$ is called the Newton's polytope of $f$.\\
\end{Def}
\begin{Def}
We set $N(f_1/f_2)=N(f_1- \Lambda f_2)$ where $\Lambda$ is a variable, and where $f_1-\Lambda f_2$ is considered as a polynomial with coefficients in $\KK[\Lambda]$.
\end{Def}
\begin{Rem}
As $\Lambda$ is a variable $N(f_1-\Lambda f_2)$ is the convex hull of $S_{f_1} \cup S_{f_2}$.
\end{Rem}
We recall the classical Bertini-Krull's theorem  in our context, see \cite[Theorem 37]{Schinzel}. 
\begin{Thm}(Bertini-Krull)\label{Bert-Krull}
Let $f_1/f_2$ a reduced rational function. Then the following conditions are equivalent:
\begin{enumerate}
\item  $f_1/f_2$ is composite,
\item 
\begin{enumerate}
\item either there exist $h_1$, $h_2 \in \overline{\KK}[\underline{X}]$ with $\deg_{\underline{X}} f_1(\underline{X})-\Lambda f_2(\underline{X})>\max(\deg h_1,\deg h_2)$ and $a_i(\Lambda) \in \overline{\KK}[\Lambda]$, such that 
$$ f_1(\underline{X})-\Lambda f_2(\underline{X}) = \sum_{i=0}^{e} a_i(\Lambda) h_1(\underline{X})^ih_2(\underline{X})^{e-i};$$
\item or the characteristic $p$ of $\KK$ is positive and $f_1(\underline{X})-\Lambda f_2(\underline{X}) \in \overline{\KK}[\Lambda][X_1^p,\dots,X_n^p]$.
\end{enumerate}
\end{enumerate}
\end{Thm}

\begin{Lem}
If $f_1/f_2$ is a composite rational function   and  the characteristic $p$ of $\KK$ is such that $p=0$ or $p>d$, then there exist $e\in \NN$, $h_1,h_2 \in \KK[\underline{X}]$ such that $N(f_1/f_2)=eN(h_1/h_2)$.
\end{Lem}

\begin{proof}
By Theorem \ref{Bert-Krull} we have $ f_1(\underline{X})-\Lambda f_2(\underline{X}) = \sum_{i=0}^{e} a_i(\Lambda) h_1(\underline{X})^i h_2(\underline{X})^{e-i}$. We denote by $u(\Lambda, \chi)$ the polynomial 
$$u(\Lambda,\chi)=\sum_{i=0}^e a_i(\Lambda)\chi^i=a_e(\Lambda)\prod_{i=1}^e\big(\chi - \varphi_i(\Lambda)\big),$$
where $\varphi_i(\Lambda) \in \overline{\KK(\Lambda)}$.\\
Thus
$$ f_1(\underline{X})-\Lambda f_2(\underline{X}) = a_e(\Lambda)\prod_{i=1}^e\Big(h_1(\underline{X}) - \varphi_i(\Lambda)h_2(\underline{X})\Big).$$
All the factors  $h_1(\underline{X}) - \varphi_i(\Lambda)h_2(\underline{X}) \in \overline{\KK(\Lambda)}[\underline{X}]$ have the same support.\\
Indeed, if we suppose the converse then there exist a coefficient $c_1 \in \overline{\KK}$ of $h_1$ and a coefficient $c_2 \in \overline{\KK}$ of $h_2$ and two indices $i$ and $j$ such that: 
$$c_1 -\varphi_i(\Lambda)c_2=0, \quad \quad c_1 -\varphi_j(\Lambda)c_2\neq 0.$$
Then $c_2 \neq 0$ and $ \varphi_i(\Lambda)=c_1/c_2 \in \overline{\KK}$. Thus $h_1 - \varphi_i(\Lambda) h_2 \in \overline{\KK}[\underline{X}]$ is a factor of $f_1(\underline{X}) -\Lambda  f_2(\underline{X})$. This implies $ f_1(\underline{X})-\Lambda f_2(\underline{X})$  is reducible in $\overline{\KK}[\Lambda][\underline{X}]$. This is impossible because $f_1$ and $f_2$ are coprime.\\
Then, for all $i=1,\dots,e$, we have:
$$N\big(h_1 - \varphi_i(\Lambda) h_2\big)=N(h_1 - \Lambda h_2)=N(h_1/h_2).$$
We recall that $F=F_1.F_2$ implies $N(F)=N(F_1)+N(F_2)$, see for example \cite[Lemma 5]{Gao_Rodrigues}, where the sum is the Minkowski's sum of  convex sets.
 Thus  we have:
$$N(f_1/f_2)= N(f_1-\Lambda f_2)=\sum_{i=1}^eN\big(h_1 - \varphi_i(\Lambda)h_2\big)=\sum_{i=1}^e N(h_1/h_2)=eN(h_1/h_2).$$
This is the desired result.
\end{proof}

The previous lemma says that if $f$ is composite then  all the vertices of $N(f)$ have a common factor: $e$. This gives our  indecomposability test designed for sparse polynomials $f_1$ and $f_2$:
\begin{Cor}[Indecomposability test]\label{test_indecomp}
Let $p$ be the characteristic of $\KK$, and $p=0$ or $p>d$.\\
Let $(i_1^{(1)},\dots,i_n^{(1)}),\dots,(i_1^{(k)},\dots,i_n^{(k)})$ be the vertices of $N(f)$.\\
If $\gcd( i_1^{(1)},\dots,i_n^{(1)},\dots,i_1^{(k)},\dots,i_n^{(k)}     )=1$ then $f$ is non-composite.
\end{Cor}

\section{Computation of a L\"uroth's generator}
In this section we show how to compute a L\"uroth's generator.  We give two algorithms.  The first one follows the strategy proposed in \cite{Sederberg} for univariate rational functions. The second one uses the algorithm \textsf{Decomp} and the computation of a greatest common right component of a univariate rational function. 
\subsection{Generalization of Sederberg's algorithm}
In this subsection, we generalize Sederberg's algorithm. Sederberg's algorithm, see \cite{Sederberg}, is a probabilistic algorithm to compute a L\"uroth's generator in the univariate case. Here, we show that the same strategy works in the multivariate case. Our algorithm is also a kind of probabilistic version of the algorithm presented in \cite{Gut_Rub_Sev}. Indeed, here we compute gcd of polynomials of the following kind $f_2(\underline{a})f_1(\underline{X}) - f_1(\underline{a})f_2(\underline{X})$, where $\underline{a} \in \KK^n$. In \cite{Gut_Rub_Sev}, the authors compute gcd of polynomials of the following kind $f_2(\underline{Y})f_1(\underline{X}) - f_1(\underline{Y})f_2(\underline{X})$, where $\underline{Y}$ are new independent variables.\\

\textbf{\textsf{Sederberg Generalized}}\\
\textbf{Input:} $f(\underline{X})=f_1/f_2(\underline{X}) $,  $g(\underline{X})=g_1/g_2(\underline{X}) \in \KK(X_1,\dots,X_n)$ two reduced rational functions, $\underline{a}$, $\underline{b} \in \KK^n$, $n\geq 2$ .\\
 \textbf{Output:} $h(\underline{X}) \in \KK(\underline{X})$ such that $\KK(f,g) =\KK(h)$, if $h$ exists.

\begin{enumerate}
\item  $F_a:=f_2(\underline{a})f_1(\underline{X}) - f_1(\underline{a})f_2(\underline{X})$, $G_a:=g_2(\underline{a})g_1(\underline{X}) - g_1(\underline{a})g_2(\underline{X})$.\\
$H_a:=\gcd(F_a,G_a)$.\\
If $H_a$ is constant then Return ``No L\"uroth's generator'', else go to \ref{Sederbergstep2}.
\item \label{Sederbergstep2}$F_b:=f_2(\underline{b})f_1(\underline{X}) - f_1(\underline{b})f_2(\underline{X})$, $G_b:=g_2(\underline{b})g_1(\underline{X}) - g_1(\underline{b})g_2(\underline{X})$.\\
$H_b:=\gcd(F_b,G_b)$.\\
If $H_b$ is constant then Return ``No L\"uroth's generator'', else go to \ref{Sederbergstep3}.
\item \label{Sederbergstep3} Return $h:=H_a/H_b$.\\
\end{enumerate}

\begin{Ex}
\begin{enumerate}
\item[a-] We set $f=X$, and $g=Y$, $\underline{a}=(0,0)$, $\underline{b}=(1,0)$. Thus $F_a=X$, $G_a=Y$ and $H_a=1$. The algorithm \textsf{Sederberg Generalized} gives $\KK(f,g)=\KK(X,Y)$ has ``No L\"uroth's generator''.
\item[b-] We consider $f=U \circ h$ and $g=V \circ h$ where $h=(X^3+Y^3+1)/(3XY)$, $U=T^2/(T+1)$, $V=(T+2)/(T^3+3)$. $h$ is a non-composite rational function.\\
We set $\underline{a}=(0,0)$, $\underline{b}=(2,1)$. In this situation we have:
$$H_a=3XY, \textrm{ and } H_b=12.(X^3+Y^3-5XY+1).$$
The algorithm \textsf{Sederberg Generalized} returns $H_a/H_b$. This is a correct output because $\KK(f,g)=\KK(h)$ and $h=u \circ (H_a/H_b)$ where $u$ is the rational function $u=(20T+1)/(12 T)$.\\
Now, if we set $\underline{a}=(0,0)$, $\underline{b}=(0,1)$ then we get $H_a=3XY$ and $H_b=12XY$. In this situation the output $H_a/H_b$ is not correct. We are in a situation where $h(\underline{a})=h(\underline{b})$ and we will see that we must avoid this situation.
\end{enumerate}
\end{Ex}

\begin{Prop}\label{Sederberg_gen_correct}
There exists an open Zariski set $U \subset \KK^{2n}$ related to $f_1$ and $f_2$, such that for all $(\underline{a}, \underline{b}) \in U$ the tree corresponding to $(\underline{a}, \underline{b})$ is executable on $f,g$ and returns (if it exists) $h$ such that $\KK(h)=\KK(f,g)$.
\end{Prop}
In order to prove this proposition we recall some results.

\begin{Def}
Given $\mathsf{f}_1,\dots,\mathsf{f}_m \in \KK(\underline{X})$, we say that they have a common right component (CRC) $h$, if there are rational functions $u_i \in \KK(T)$, $i=1,\dots, m$,  such that $\mathsf{f}_i=u_i \circ h$, and $\deg u_i >1$.\\
 $h$ is a greatest common right component (GCRC) of $\mathsf{f}_1,\dots,\mathsf{f}_m$ if the $u_i's$ have not a common right component of degree greater than one.
\end{Def}

\begin{Prop}\label{GCRC_prop}
$\KK(\mathsf{f}_1,\dots,\mathsf{f}_m)=\KK(h)$ if and only if $h$ is a GCRC of $\mathsf{f}_1$,\dots,$\mathsf{f}_m$.
\end{Prop}
\begin{proof}
This proposition is proved in the univariate case in \cite{Alonso_Gut_Recio} but the proof can be extended to the multivariate case in a straightforward way.
\end{proof}

\begin{proof}[Proof of Proposition \ref{Sederberg_gen_correct}]
Firstly,  we suppose that there exists a L\"uroth's generator $h=h_1/h_2$, where $h_1/h_2$ is reduced. Then, by Proposition \ref{GCRC_prop}, $f=u \circ h$ and $g=v \circ h$ where $u,v \in \KK(T)$ do not have a common right component of degree greater than one. Thus $\KK\big(u(T),v(T)\big)=\KK(T)$.   Then there exist $Q_1,Q_2 \in \KK[U,V]$ such that $Q_1\big( u(T),v(T) \big) / Q_2\big( u(T),v(T) \big) =T$.\\
Furthermore by Lemma \ref{Lem_Bodin}, 
$$F_a=f_2(\underline{a})f_1(\underline{X}) -f_1(\underline{a})f_2(\underline{X})=e \prod_i\big( h_1(\underline{X})-t_i h_2(\underline{X}) \big)$$ where $e \in \KK$ and $t_i$ are the roots of 
$$f_2(\underline{a})u_1(T) -f_1(\underline{a})u_2(T)=:u_a,$$
and
$$G_a=g_2(\underline{a})g_1(\underline{X}) -g_1(\underline{a})g_2(\underline{X})=e' \prod_i\big( h_1(\underline{X})-s_i h_2(\underline{X}) \big)$$  
 where $e' \in \KK$ and $s_i$ are the roots of 
$$g_2(\underline{a})v_1(T) -g_1(\underline{a})v_2(T)=:v_a.$$
We get: $h(\underline{a})$ is a common root of $u_a$ and $v_a$. Thus $h_1(\underline{X}) -h(\underline{a})h_2(\underline{X})$ divides $F_a$ and $G_a$.\\
If $f_2(\underline{a}).g_2(\underline{a}).Q_2\big(u(h(\underline{a})),v(h(\underline{a}))\big) \neq 0$ then $h(\underline{a})$ is the unique common root of $u_a$ and $v_a$. Indeed if there exists another root $x$ such that $u_a(x)=v_a(x)=0$, then $u\big(h(\underline{a})\big)=f_1(\underline{a})/f_2(\underline{a})=u(x)$ and $v \big( h(\underline{a})\big)=g_1(\underline{a})/g_2(\underline{a})=v(x)$.\\
It follows:
$$h(\underline{a})= \dfrac{Q_1\big( u(h(\underline{a})),v(h(\underline{a}))\big)}{ Q_2\big(u(h(\underline{a})),v(h(\underline{a}))\big)}=\dfrac{Q_1\big(u(x),v(x)\big)}{ Q_2\big(u(x),v(x)\big)}=x.$$
Now we remark that if $t  \neq s$ then $\gcd(h_1+th_2, h_1+sh_2)$ is constant.\\
We get then: $\gcd(F_a,G_a)=h_1(\underline{X}) -h(\underline{a})h_2(\underline{X})$.\\
In the same way:  $\gcd(F_b,G_b)=h_1(\underline{X}) -h(\underline{b})h_2(\underline{X})$.\\
If $h(\underline{a}) \neq h(\underline{b})$, this gives the desired result, because $\KK(h)=\KK(H)$ when $H=U \circ h$ with $U=\big(T-h(\underline{a})\big)/ \big( T - h(\underline{b})\big)$.\\

Secondly, we suppose that there does not exist a L\"uroth's generator.\\
Then we have  $f=u \circ h$ and $g=v \circ H$, with $h, H \in\KK(\underline{X})$ non-composite and algebraically independent.\\
Thus $F_a(\underline{X})=e. \prod_i\big( h_1(\underline{X})-t_i h_2(\underline{X}) \big)$ as before, with $h_1(\underline{X}) -t_i h_2(\underline{X})$ absolutely irreducible if $t_i \not \in \sigma(h_1,h_2)$. The condition $t_i \not \in \sigma(h_1,h_2)$ means 
$$R(\underline{a})=Res_T\big( f_2(\underline{a})u_1(T)-f_1(\underline{a})u_2(T), \, \Spect_{h_1,h_2}(T)\big)  \neq 0,$$
 where $\Spect_{h_1,h_2}(T)=\prod_{\lambda \in \sigma(h_1,h_2) \cap \KK}(T- \lambda)$.\\
In the same way, we have $G_a=e'.\prod_i(H_1(\underline{X}) -s_i H_2(\underline{X}))$ with\\ $H_1(\underline{X}) -s_i H_2(\underline{X})$ absolutely irreducible if 
$$S(\underline{a}) =Res_T \big( g_2(\underline{a})v_1(T) -g_1(\underline{a})v_2(T),\, \Spect_{H_1,H_2}(T)\big) \neq 0.$$
Thus $F_a$ and $G_a$ have a non trivial common divisor if and only if there exist $t_i$, $s_j$ and $\alpha \in \KK \setminus \{0\}$ such that:
$$(\star)\, \alpha\big( h_1(\underline{X}) - t_i h_2(\underline{X}) \big)= H_1(\underline{X}) -s_j H_2(\underline{X}).$$
In the same way, $F_b$ and $G_b$ have  a non trivial common divisor if and only if  there exists $t'_i$, $s'_j$ and $\alpha' \in \KK \setminus \{0\}$ such that: 
$$(\star\star)\,\alpha'\big( h_1(\underline{X}) - t'_i h_2(\underline{X}) \big)= H_1(\underline{X}) -s'_j H_2(\underline{X}).$$
$(\star)$ and $(\star \star)$ give:
$$\begin{pmatrix}
\alpha & -\alpha t_i\\
\alpha' & -\alpha' t'_i 
\end{pmatrix}
\begin{pmatrix}
h_1\\
h_2\\
\end{pmatrix}
=
\begin{pmatrix}
1 & -s_j\\
1& -s'_j 
\end{pmatrix}
\begin{pmatrix}
H_1\\
H_2\\
\end{pmatrix}.
$$
If $$D(\underline{a}, \underline{b})=Res_T \big( g_2(\underline{a})v_1(T)-g_1(\underline{a})v_2(T), \, g_2(\underline{b})v_1(T)-g_1(\underline{b})v_2(T)\big) \neq 0$$
 then $s_j \neq s'_j$ and the previous system gives $H=u \circ h$, with $\deg u =1$.
 Thus $h$ and $H$ are algebraically dependent and this is absurd. Thus $F_a$ and $G_a$ (resp. $F_b$ and $G_b$) have no common divisor.\\
Hence, if no L\"uroth's generator exists and $f_2(\underline{a}).g_2(\underline{b}).R(\underline{a}).S(\underline{a}).R(\underline{b}).S(\underline{b}).D(\underline{a},\underline{b})$ is not equal to zero, then $\gcd( F_a,G_a)$ is constant and $\gcd( F_b,G_b)$ is constant. Thus the algorithm returns ``No L\"uroth's generator''.
\end{proof}

\begin{Rem}\label{remarkzariskiluroth1}
With the notations of the previous proof, we remark that $\underline{a}$ and $\underline{b}$ must avoid the roots of: $f_2(\underline{X})$, $g_2(\underline{X})$, $h_2(\underline{X})$, $Q_2\big( f(\underline{X})\big),g(\underline{X})\big)$, $R(\underline{X})$, $S(\underline{X})$, and $(\underline{a}, \underline{b})$ must avoid the roots of $h_1(\underline{A})h_2(\underline{B})-h_1(\underline{B})h_2(\underline{A})$ and $D(\underline{A},\underline{B})$.\\
We can easily bound  the degree of each polynomial: $\deg f_i \leq d$, $\deg g_i \leq d$, $\deg h_i \leq d/2$, $\deg Q_2 \leq d(d-1)$ see \cite[Proposition 2.1]{BuseAndrea},  $\deg R \leq d(d^2-1)$, $\deg S \leq d(d^2-1)$, and $\deg D \leq d^3$.\\
Then if $\KK$ is ``big enough'' the open Zariski set $U$ is not the empty set.
\end{Rem}

\begin{Rem}
In the algorithm \textsf{Sederberg Generalized} we cannot consider two random linear combinations of $f_1$, $f_2$ and $g_1$, $g_2$. Indeed,  with random linear combinations and with the notations of the previous proof, $u_a$ and $v_a$  do not have a unique common root in $\KK$. Thus with random linear combinations the strategy used in Proposition \ref{Sederberg_gen_correct} is not  valid.
\end{Rem}

\begin{Prop}\label{sederberg_complexity}
If $\KK$ is a field with at least $(4d+2)d$ elements then the algorithm \textsf{Sederberg Generalized} uses $\bigOsoft(d^n)$ arithmetic operations. 
\end{Prop}

\begin{proof}
The computations of $f_i(\underline{a})$, $g_i(\underline{a})$, $f_i(\underline{b})$, $g_i(\underline{b})$ needs $\bigOsoft(d^n)$ arithmetic operations. The complexity of an $n$-variate gcd computation needs $\bigOsoft(d^n)$ arithmetic operations. Indeed, as $\KK$ is a field with at least $(4d+2)d$ elements with Lemma 6.44 in \cite{GG} we can generalize to $n$ variables the algorithm 6.36 presented in \cite{GG} and obtain a result like Corollary 11.9 in \cite{GG}. This gives the desired result.
\end{proof}

\begin{Rem}
When it is possible, a polynomial generator is desirable. The algorithm \textsf{Sederberg Generalized} always returns a rational generator. We can test if we have a polynomial generator in the following way: We test if there exist $\alpha, \beta \in \KK$ such that $\alpha H_a +\beta=H_b$. If such constants exist then $H_a$ (or $H_b$) is a polynomial generator.\\
This improvement is correct because we have seen during the proof of Proposition \ref{Sederberg_gen_correct} that $H_a=h_1-h(\underline{a})h_2$ and $H_b=h_1-h(\underline{b})h_2$. Thus if a polynomial generator $h_1$ exists we have $H_a=h_1-h_1(\underline{a})$ and $H_b=h_1-h_1(\underline{b})$. As gcd are known up to a multiplicative constant there exist  $\alpha, \beta \in \KK$ such that $\alpha H_a+\beta=H_b$. Conversely, if we have $\alpha H_a+\beta=H_b$ then $H_a/H_b=u \circ H_a$ with $u=T/\big(\alpha T + \beta\big)$, thus $\KK(H_a/H_b)=\KK(H_a)$.\\
The computation of $\alpha$ and $\beta$ needs $\bigO(d^n)$ arithmetic operations. Indeed, we solve a linear system with $\bigO(d^n)$ equations and two unknowns. Thus we can find a polynomial generator with the algorithm \textsf{Sederberg Generalized} with $\bigOsoft(d^n)$ arithmetic operations.
\end{Rem}

\subsection{Another strategy based on decomposition}
Now, we give another  algorithm to compute a L\"uroth's generator. Here we use the relation between decomposition and computation of a L\"uroth's generator. \\

\textbf{\textsf{L\"uroth with Decomp}}\\
\textbf{Input:} $f(\underline{X})=f_1/f_2(\underline{X}) $,  $g(\underline{X})=g_1/g_2(\underline{X}) \in \KK(X_1,\dots,X_n)$ two reduced rational functions, $\underline{z}:=(\underline{a},\underline{b})\in \KK^{2n}$.\\
 \textbf{Output:} $h(\underline{X}) \in \KK(\underline{X})$ such that $\KK(f,g) =\KK(h)$, if $h$ exists.

\begin{enumerate}
\item  Decompose $f$ with the algorithm \textsf{Decomp}, then $f=u \circ h$.
\item Compute $v$ (if it exists) such that $g=v \circ h$.
\item  If $v$ do not exist then Return ``No L\"uroth's generator'', else go to \ref{Luroth_decompstep3}.
\item \label{Luroth_decompstep3} Compute $w$ the GCRC of $u$ and $v$ with Sederberg's algorithm.
\item Return $w\circ h$.
\end{enumerate}

\begin{Prop}
The algorithm \textsf{L\"uroth's with Decomp} is correct for $z $ satisfying the hypothesis of Theorem \ref{main_thm}.
\end{Prop}

\begin{proof}
 This algorithm computes a GCRC of $f$ and $g$, thus by Proposition  \ref{GCRC_prop}, this gives the desired result.
\end{proof}

\begin{Prop}\label{luroth_decomp_complexity}
Under hypotheses (C) and (H), the algorithm \textsf{L\"uroth's with Decomp} performs one factorization of a univariate polynomial of degree $d$ over $\KK$ plus a number of operations in $\KK$ belonging to $\bigOsoft(d^n)$ if $n\geq 3$ or to $\bigOsoft(d^{3})$ if $n=2$. 
\end{Prop}

\begin{proof}
The first step of the algorithm performs one factorization of a univariate polynomial of degree $d$ over $\KK$ plus a number of operations in $\KK$ belonging to $\bigOsoft(d^n)$ if $n\geq 3$ or to $\bigOsoft(d^{3})$ if $n=2$ by Proposition \ref{prop_complexity}.\\
With the strategy presented in Section \ref{sec_comput_u}, the second step can be done with $\bigOsoft(d^{n})$ arithmetic operations.\\
The last step can be done in an efficient probabilistic way,  see \cite{Sederberg}. The algorithm presented in \cite{Sederberg} computes only two gcd's of univariate polynomials of degree lower than $d$.\\
Then the total cost of the algorithm belongs to $\bigOsoft(d^n)$ if $n\geq 3$ or to $\bigOsoft(d^{3})$ if $n=2$.
\end{proof}

\begin{Rem}\label{remarkzariskiluroth2}
During the algorithm \textsf{L\"uroth with Decomp} we have to avoid the roots of  nonzero polynomials considered in Remark \ref{remark_poly_sans_hyp_complexity} and Remark \ref{remarksomepoly} because we use the algorithm \textsf{Decomp}. Furthermore during the algorithm \textsf{L\"uroth with Decomp}, we use Sederberg's algorithm, this algorithm is also probabilistic and has in input two parameters $x_1, x_2 \in \KK$. If $x_1$ and $x_2$ are not the roots of a nonzero polynomials then the output is correct, see \cite{Sederberg}.\\
Thus the nonzero polynomials are just the ones used for the algorithm \textsf{Decomp} and for  Sederberg's algorithm. 
\end{Rem}

\subsection{Computation of a L\"uroth's generator}
$\,$\\
\textbf{\textsf{L\"uroth's generator}}\\
\textbf{Input:} $\mathsf{f}_1(\underline{X}),\dots,\mathsf{f}_m(\underline{X}) \in \KK(\underline{X}) $,  $m$ reduced rational functions,\\
$z:=\underline{z}_2,\dots,\underline{z}_m \in \KK^{2n}$, $n\geq 2$.\\
 \textbf{Output:} $h(\underline{X}) \in \KK(\underline{X})$ such that $\KK(\mathsf{f}_1,\dots,\mathsf{f}_m) =\KK(h)$, if $h$ exists.

\begin{enumerate}
\item  Compute a L\"uroth's generator of $\KK(\mathsf{f}_1,\mathsf{f}_2)$ with \textsf{Sederberg Generalized} applied to $\mathsf{f}_1$, $\mathsf{f}_2$, with $z_2$ .
\item If a L\"uroth's generator $h$ is found then go to step \ref{luroth_gen_step3} else Return ``No L\"uroth's generator''. 
\item\label{luroth_gen_step3} For $i=3,\dots,m$,
\begin{enumerate}
\item  Compute a L\"uroth's generator of $\KK(h,\mathsf{f}_i)$ with \textsf{Sederberg Generalized} applied to $h$, $\mathsf{f}_i$, with $z_i$.
\item If a L\"uroth's generator $H$ is found then $h:=H$ else Return ``No L\"uroth's generator''. 
\end{enumerate}
\item Return h.
\end{enumerate}

\begin{Prop}\label{mlurothcorrect}
The algorithm \textsf{L\"uroth's generator} is correct for $z$ satisfying the hypothesis of Theorem \ref{mainluroth}.
\end{Prop}

\begin{proof}
We just have to remark that $\KK(\mathsf{f}_1,\dots,\mathsf{f}_{i-1},\mathsf{f}_i)=\KK(\mathsf{f}_1,\dots,\mathsf{f}_{i-1})(\mathsf{f}_i)$. 
\end{proof}

\begin{Prop}\label{mlurothcomplexity}
If $\KK$ has at least $(4d+2)d$ elements, then the algorithm \textsf{L\"uroth's generator} can be performed with $\bigOsoft(md^n)$ arithmetic operations in $\KK$.
\end{Prop}

\begin{proof}
We use  $m$ times  the algorithm  \textsf{Sederberg Generalized}. Thus, thanks to Proposition \ref{sederberg_complexity}  we get the desired complexity.
\end{proof}

\begin{Rem}
During the algorithm \textsf{L\"uroth's generator}  we can use the algorithm \textsf{L\"uroth with Decomp} instead of \textsf{Sederberg Generalized}. In the bivariate case, the complexity becomes then $\bigOsoft(d^{3})$. In this case the algorithm is not softly optimal, but the algorithm can also return $u$ such that $f=u \circ h$.
\end{Rem}

We conclude that Proposition \ref{mlurothcorrect} and Proposition \ref{mlurothcomplexity} prove Theorem \ref{mainluroth}.

\section{Study of the Gutierez-Rubio-Sevilla's algorithm}\label{appendix}
In this section we study the complexity of the decomposition algorithm given in \cite{Gut_Rub_Sev}.  More precisely, we explain how to modify it in order to get a polynomial time algorithm instead of an exponential time algorithm. 
\subsection{Some preliminary results}
The following lemma is a generalization of Lemma \ref{Lem_Bodin}. 
\begin{Lem}\label{Lem_Bodin_gen}
Let $h=h_1/h_2$ be a  rational function in $\KK(\underline{X})$, $u=u_1/u_2$ a rational function in $\KK(T)$ and set $f=u \circ h$ with $f=f_1/f_2 \in \KK(\underline{X})$. 
Let $\lambda, \mu \in \LL$, where $\LL$ is a field and $\KK \subset \LL$. We have:
$$\mu f_1 - \lambda f_2= (\mu u_1 - \lambda u_2)(h).h_2^{\deg u}. $$
\end{Lem}
\begin{proof}
We have $$\dfrac{\mu f_1- \lambda f_2}{f_2}=\mu\dfrac{u_1(h)}{u_2(h)}- \lambda\dfrac{u_2(h)}{u_2(h)}=\dfrac{\mu u_1(h) - \lambda u_2(h)}{u_2(h)}.$$
Thus: $(\star) \, \,(\mu f_1 - \lambda f_2).u_2(h)=(\mu u_1- \lambda u_2)(h).f_2$.\\
Furthermore 
$$(\star \star) \, \, \dfrac{f_1}{f_2}=\dfrac{u_1(h)}{u_2(h)} =\dfrac{ \big( \sum_{i=0}^{d_1} a_i h_1^ih_2^{d_1-i} \big). h_2^{d_2} }{  \big( \sum_{i=0}^{d_2} b_i h_1^ih_2^{d_2-i} \big). h_2^{d_1}     },$$
 where $u_1(T)=\sum_{i=0}^{d_1}a_iT^i$, $u_2(T)=\sum_{i=0}^{d_2}b_i T^i.$\\
Then $f_2=\big( \sum_{i=0}^{d_2} b_ih_1^ih_2^{d_2-i} \big). h_2^{\max(d_1-d_2,0)}$ because $f$ is reduced and the degree of the right  term of $(\star \star)$ is lower or equal to $\deg(f)$.\\
It follows $f_2=u_2(h).h_2^{\max(d_1-d_2,0)+d_2}=u_2(h).h_2^{\deg u}$, then thanks to $(\star)$ we deduce the desired result.
\end{proof}

\begin{Prop}\label{h=wphi}
Let $f\in \KK(\underline{X})$ be a  rational function such that $f=u \circ h$ and $f=u \circ \varphi$, where $u$ is a  rational function in $\KK(T)$, $h$  a  non-composite rational function and $\varphi$ a  rational function.\\
Then $\varphi$ is non-composite and there exists $w \in \KK(T)$ such that $h= w \circ \varphi$ and $\deg w=1$.
\end{Prop}

\begin{Rem}
$w$ is not necessarily the identity. For example if $u=x^2+1/x^2$ and $w=1/x$ then $u \circ w=u$. Thus we can get $f=(u \circ w) \circ \varphi=u \circ \varphi$ and $f=u\circ (w \circ \varphi)=u \circ h$. See \cite{Gut_Sev_contre_exemple} for more statements on the particular situation $u \circ w=u$.
\end{Rem}

\begin{proof}
We set $u=u_1/u_2$ and $\varphi=\varphi_1/\varphi_2$.\\
Let $\lambda, \mu \in \overline{\KK}$ such that $\deg (\mu u_1 - \lambda u_2 )=\deg u$, by Lemma \ref{Lem_Bodin_gen} we have
$$\mu f_1 - \lambda f_2 = e\prod_{i=1}^{\deg u} (h_1 -x_i h_2),$$
where $e \in \overline{\KK}$ and $x_i \in \overline{\KK}$ are the roots of $\mu u_1 - \lambda u_2$ .\\
We can  suppose that $h_1 -x_i h_2$ are absolutely irreducible and $x_i \neq x_j$ if $i \neq j$.\\
Indeed, the ``bad'' values of $(\mu: \lambda)$ are $(u_2(x):u_1(x))$ where $x \in \sigma(h_1,h_2)$ and are the roots of $R(\mu,\lambda)=Res(\mu u_1- \lambda u_2,\mu u'_1 - \lambda u'_2)$. As $\sigma(h_1,h_2)$ is finite and $\overline{\KK}$ infinite, we deduce that ``good'' values of $(\mu: \lambda)$ exist.\\
We can also suppose that $\deg \varphi_1 -x_i \varphi_2=\deg \varphi$, because we just have to avoid a finite number of $x_i$.\\
Then Lemma \ref{Lem_Bodin_gen} also implies 
$$\mu f_1 - \lambda f_2 = e\prod_{i=1}^{\deg u} (\varphi_1 -x_i \varphi_2).$$
We have $\varphi_1 -x_i \varphi_2$ is absolutely irreducible, else $\mu f_1 - \lambda f_2$ has more than $\deg u$ absolute irreducible factors: this is a contradiction with $h_1 -x_i h_2$ being absolutely irreducible.\\
Then  $\varphi$ is non-composite by  Proposition \ref{Thm_Bodin}.\\
Furthermore, there exist $i_k,j_k$, with $k=1,\dots,\deg u$ such that $h_1 -x_{i_k}h_2$ equal $\varphi_1 -x_{j_k}\varphi_2$ up to a multiplicative constant. As in the proof of Proposition \ref{Sederberg_gen_correct} it follows $\varphi=w\circ h$ with $w \in \overline{\KK}(T)$ and $\deg w=1$. As $h$ and $\varphi$ belongs to $\KK(\underline{X})$ we have $w \in \KK(T)$. (Indeed we just have to solve a linear system in $\KK$ to get $w$.)
\end{proof}

\subsection{Study of the absolute irreducible factors of near-separated polynomials}
The decomposition algorithm given in \cite{Gut_Rub_Sev} is based on the following theorem; see \cite{Schicho}. In this subsection we improve this result.
\begin{Thm}\label{schichothm}
Let $f=f_1/f_2 \in \KK(\underline{X})$.\\
$f=u \circ h$, with $h=h_1/h_2$    if and only if $H(\underline{X},\underline{Y})=h_1(\underline{X})h_2(\underline{Y})- h_2(\underline{X})h_1(\underline{Y})$ divides $F(\underline{X},\underline{Y})=f_1(\underline{X})f_2(\underline{Y})- f_2(\underline{X})f_1(\underline{Y})$.
\end{Thm}

In the following we use a result due to Schinzel.
\begin{Def}
A rational function is reducible over $\KK$ if the numerator in its reduced form is reducible over $\KK$.
\end{Def}

\begin{Lem}\label{lemschinzel}
Let $\Psi(T,\underline{Y})$ and $f(\underline{X})$ be non-constant rational functions over $ \KK$, the former  of non-negative degree  with respect to $T$  and to at least one $Y_i$.\\
If the function $$\psi\big(f(\underline{X}),\underline{Y}\big)$$
is reducible over $\KK$ then 
$f=u\circ h, \, u \in \KK(T), \, h \in \KK(\underline{X})$ and $\psi\big(u(T), \underline{Y}\big)$ is reducible over $\KK$.
 \end{Lem}

\begin{proof}
See \cite[Lemma 1]{Sc1985}.
\end{proof}

\begin{Prop}\label{appendixprop}
Let $f =f_1/f_2 \in \KK(\underline{X})$, $\hat{f} =\hat{f}_1/\hat{f}_2 \in \KK(\underline{Y})$ be  two  non-constant rational functions.\\
If $f$ and $\hat{f}$ are  non-composite  then $F(\underline{X},\underline{Y})=f_1(\underline{X})\hat{f}_2(\underline{Y})- f_2(\underline{X})\hat{f}_1(\underline{Y})$ is irreducible in $\KK[\underline{X},\underline{Y}]$.
\end{Prop}

\begin{proof}
We set $\psi(T,\underline{Y})=\hat{f}(\underline{Y})-T$.\\
Then $$\psi\big( f(\underline{X}),\underline{Y} \big)= \dfrac{\hat{f}_1(\underline{Y})f_2(\underline{X}) -f_1(\underline{X})\hat{f}_2(\underline{Y})}{f_2(\underline{X})\hat{f}_2(\underline{Y})}.$$
If we suppose $F( \underline{X}, \underline{Y})$ reducible then $f=u \circ h$ and $\psi\big( u(T), \underline{Y})$ is reducible by Lemma \ref{lemschinzel}.\\
As $f$ is non-composite $\deg u =1$ thus we can set $u(T)=(aT+b)/(\alpha T + \beta)$. Then $\psi\big(u(T), \underline{Y})$ is reducible means $\hat{f}_1(\underline{Y})(\alpha T+ \beta) - \hat{f}_2(\underline{Y})(aT+b)$ is reducible over $\KK$. By Proposition \ref{Thm_Bodin} this is absurd because $\hat{f}$ is non-composite. Hence $F(\underline{X}, \underline{Y})$ is irreducible.
\end{proof}

Now we can improve Theorem \ref{schichothm}.

\begin{Thm}\label{schicho_mieux}
Let $f=f_1/f_2 \in \KK(\underline{X})$ a non-constant  rational function.\\
If $f=u \circ h$, where $u=u_1/u_2 \in \KK(T)$ and $h=h_1/h_2 \in \KK(\underline{X})$ are  rational functions, with $\deg u \geq 2$ and $h$  non-composite, then the irreducible factors with the smallest degree relatively to $\underline{X}$ of 
$$F(\underline{X},\underline{Y})=f_1(\underline{X})f_2(\underline{Y})-f_2(\underline{X})f_1(\underline{Y})$$
are of the kind
$$H(X,Y)=h_1(\underline{X})\varphi_{i,2}(\underline{Y})-h_2(\underline{X})\varphi_{i,1}(\underline{Y}),$$
where $\varphi_i=\varphi_{i,1}/\varphi_{i,2}$ are  non-composite rational functions such that \mbox{$h=w\circ \varphi_i$} with $\deg w=1$.
\end{Thm}

Theorem \ref{mainschichothm} is a direct consequence of Theorem \ref{schicho_mieux}.

\begin{proof}
By Lemma \ref{Lem_Bodin_gen}, we have 
$$(\star)\, \, F(\underline{X},\underline{Y})=U_{f_1,f_2}\big(h(\underline{X})\big).h_2(\underline{X})^{\deg u},$$ 
where 
$$U_{f_1,f_2}(T)=f_2(\underline{Y})u_1(T)-f_1(\underline{Y})u_2(T).$$
As $f=u \circ h$, $h(Y)$ is a root of $U_{f_1,f_2}$. Then 
$$U_{f_1,f_2}(T)=\big( h_2(\underline{Y})T-h_1(\underline{Y}) \big)A(\underline{Y},T),$$
 where $A(\underline{Y},T) \in \KK[\underline{Y},T]$. Thus $(\star)$  implies $h_1(\underline{X})h_2(\underline{Y})-h_2(\underline{X})h_1(\underline{Y})$ divides $F(\underline{X},\underline{Y})$.\\
Now, we suppose that $\varphi(\underline{Y}) \in \KK(\underline{Y})$ is another root of $U_{f_1,f_2}(T)$. Then 
$$u\big( \varphi(\underline{Y})\big)=f(\underline{Y})=u\big(h(\underline{Y})\big).$$
 Thus, by Lemma \ref{h=wphi}, we have $\varphi$ is non-composite and $h =w \circ \varphi$ with $\deg w=1$. As before, we can write
$U_{f_1,f_2}=\big( \varphi_2(\underline{Y})T- \varphi_1(\underline{Y}) \big).B(\underline{Y},T)$, where $B(\underline{Y},T) \in \KK[\underline{Y},T]$.\\
Thus $\varphi_2(\underline{Y})h_1(\underline{X})-\varphi_1(\underline{Y})h_2(\underline{X})$ divides $F(\underline{X}, \underline{Y})$ by $(\star)$.\\
Now, we write
$$(\star \star) \, \, U_{f_1,f_2}(T)=\prod_{i \in I} \big( \varphi_{i,2}(\underline{Y})T- \varphi_{i,1}(\underline{Y}) \big) . \prod_{j \in J}C_j^{e_j}(\underline{Y},T),$$
where $\varphi_i=\varphi_{i,1}/\varphi_{i,2}(\underline{Y})$ is a reduced non-composite rational function as explained above and $C_j(\underline{Y},T) \in\KK[\underline{Y},T]$ is irreducible with $\deg_T C_j \geq 2$.\\
We evaluate $T$ to $h$ in $(\star \star)$ and multiply the result  by $h_2^{\deg u}$:
\begin{eqnarray*}
U_{f_1,f_2}\big( h(\underline{X})\big).h_2(\underline{X})^{\deg u}&=&
\prod_{i \in I} \big( \varphi_{i,2}(\underline{Y})h_1(\underline{X})- \varphi_{i,1}(\underline{Y})h_2(\underline{X}) \big) \\
&&\times \Big(\prod_{j \in J}C_j^{e_j}\big(\underline{Y},h(\underline{X})\big)\Big).h_2(\underline{X})^{\sum_{j \in J}e_j \deg_T C_j}.
\end{eqnarray*}
The factors $\varphi_{i,2}(\underline{Y})h_1(\underline{X})- \varphi_{i,1}(\underline{Y})h_2(\underline{X})$ are irreducible by Proposition \ref{appendixprop}. Furthermore, by Lemma \ref{lemschinzel} as $h$ is non-composite and $C_j(\underline{Y},T)$ is irreducible, we have \mbox{$C_j\big( \underline{Y},h(\underline{X})\big).h_2(\underline{X})^{\deg_T C_j}$} is irreducible in $\KK[\underline{X}, \underline{Y}]$.\\
We also have 
\begin{eqnarray*}
\deg_X C_j\big( \underline{Y},h(\underline{X})\big)h_2(\underline{X})^{\deg_T C_j}&=&\deg_T C_j. \deg h \\
& \geq& 2 \deg h\\
&>& \deg_X \varphi_{i,2}(\underline{Y})h_1(\underline{X})- \varphi_{i,1}(\underline{Y})h_2(\underline{X}).
\end{eqnarray*}
Then  $H(X,Y)=\varphi_{i,2}(\underline{Y})h_1(\underline{X})- \varphi_{i,1}(\underline{Y})h_2(\underline{X})$ are the factors with the smallest degree relatively to $\underline{X}$.
\end{proof}

\subsection{Improvement of the GRS algorithm}
Now we describe the decomposition algorithm  presented in \cite{Gut_Rub_Sev}.\\

\textbf{\textsf{GRS decomposition algorithm}}\\
\textbf{Input:} $f(\underline{X})=f_1/f_2(\underline{X}) $, $n\geq 2$.\\
 \textbf{Output:} $u \in \KK(T)$, $h(\underline{X}) \in \KK(\underline{X})$   such that $f=u \circ h$, or ``$f$ is non-composite''.

\begin{enumerate}
\item \label{GRSstep1}Factor $F(\underline{X},\underline{Y})$. Let $D= \{H_1,\dots,H_m\}$ be the set of factors of $F$ (up to product by constants).  We set $i=1$.
\item \label{GRSstep2} If $H_i$ can be written $H_i(\underline{X},\underline{Y})=h_1(\underline{X})h_2(\underline{Y})-h_1(\underline{Y})h_2(\underline{X})$ then $h_1/h_2$ is a right component for $f$. Then compute $u$ by solving a linear system and Return $u$, $h$.
\item If $i<m$ then $i:=i+1$ and go to step \ref{GRSstep2}, else Return ``$f$ is non-composite''.
\end{enumerate}

This algorithm has an exponential time complexity. Indeed, the set $D$ contains at most $2^d$ polynomials, where $d$ is the degree of $f$.

However, we can improve this algorithm. Thanks to Proposition \ref{appendixprop}, we remark that if $f$ is non-composite then $F$ is irreducible. Furthermore, if $f=u \circ h$ with $h$ non-composite, then $H(X,Y)=h_1(\underline{X})h_2(\underline{Y})-h_1(\underline{X})h_2(\underline{Y})$ is an irreducible factor of $F(\underline{X},\underline{Y})$, by Theorem \ref{schicho_mieux}. Thus we have to study at most $\deg F$ irreducible factors. Thus we can substitute the set $D$ by the set of \emph{irreducible} factors. (We can also  substitute the set $D$ by the set of \emph{irreducible factors with the smallest degree relatively to $\underline{X}$}). As Step \ref{GRSstep1} and Step \ref{GRSstep2} can be done in a polynomial time, it follows:
\begin{Prop}  If in the \textsf{GRS decomposition algorithm} we set: ``$D$ is the set of irreducible factors of $F$'', then this modified algorithm has a polynomial  time complexity.
\end{Prop}

\begin{Rem}
The bottleneck of this modified algorithm is the factorization of $F$. If we apply the deterministic algorithm proposed in \cite{Lec2007} then the modified GRS decomposition algorithm uses $\bigOsoft(d^{2n+\omega-1})$ arithmetic operations, where $d$ is the degree of $f$ and $n$ the number of variables.
\end{Rem}

\begin{Ex}
Now, we illustrate the \textsf{GRS decomposition algorithm} with $f=u \circ h$, where $u=(T^2+1)/T$, $h=h_1/h_2$, and $h_1=X_1^3+X_2^3+1$, $h_2=3X_1X_2$. $h$ is a non-composite rational function.\\
In this situation, we have the following factorization of $F(X_1,X_2,Y_1,Y_2)$:
$$F(X_1,X_2,Y_1,Y_2)=3.H_1(X_1,X_2,Y_1,Y_2).H_2(X_1,X_2,Y_1,Y_2), \textrm{ where}$$
\begin{eqnarray*}
H_1(X_1,X_2,Y_1,Y_2)&=&X_1^3Y_1Y_2+X_2^3Y_1Y_2+Y_1Y_2-Y_1^3X_1X_2-Y_2^3X_1X_2-X_1X_2\\
&=&h_1(X_1,X_2)h_2(Y_1,Y_2)-h_1(Y_1,Y_2)h_2(X_1,X_2),\\
H_2(X_1,X_2,Y_1,Y_2)&=&1+X_1^3+X_2^3+Y_1^3+Y_2^3+X_1^3Y_1^3+X_1^3Y_2^3+X_2^3Y_1^3+X_2^3Y_2^3\\
&&-9X_1X_2Y_1Y_2\\
&=&h_1(X_1,X_2)h_1(Y_1,Y_2)-h_2(Y_1,Y_2)h_2(X_1,X_2).
\end{eqnarray*}
Then we can recover the decomposition $f= u \circ h$ with the \textsf{GRS decomposition algorithm}.
\end{Ex}






\end{document}
\endinput